# Investigating factors affecting learner's perception toward online learning: Evidence from ClassStart Application in Thailand

Nattaporn Thongsri[a], Liang Shen[b] and Yukun Bao [a]

[a]Center for Modern Information Management, School of Management, Huazhong University of  Science and Technology, Wuhan, People's Republic of China;
[b]Center for Big Data Analytics, Jiangxi University of Engineering, Xinyu, People's Republic of China

**ABSTRACT**

Twenty-First Century Education is a design of instructional culture that empowers learner-centered through the philosophy of "Less teaching but more learning". Due to the development of technology enhance learning in developing countries such as Thailand, online learning is rapidly growing in the electronic learning market. ClassStart is a learning management system developed to support Thailand's educational management and to promote the student-centred learning processes. It also allows the instructor to analyse individual learners through system-generated activities. The study of online learning acceptance is primarily required to successfully achieve online learning system development. However, the behavioural intention of students to use online learning systems has not been well examined, in particular, by focusing specific but representative applications such as ClassStart in this study. This research takes the usage of ClassStart as research scenario and investigates the individual acceptance of technology through the Unified Theory of Acceptance and Use of Technology, as well as technological quality through the Delone and McLean IS success model. A total of 307 undergraduate students using ClassStart responded to the survey. The Partial Least Squares method, a statistics analysis technique based on the Structural Equation Model (SEM), was used to analyze the data. It was found that performance expectancy, social influence, information quality and system quality have the significant effect on intention to use ClassStart.

* Corresponding author: Tel: +86-27-87558579; fax: +86-27-87556437.
Email: yukunbao@ hust.edu.cn or y.bao@ieee.org

## 1. Introduction

Classrooms' walls are destroyed with information technology (Al-Mushasha & Nassuora, 2012). A role of the internet is that it eliminates the time gap between learners and instructors (Kibelloh & Bao, 2014a). The quick evolution of IT results in new applications that can meet people's needs in various terms such as e-banking, e-commerce, e-health and e-learning (Alsabawy, Cater-Steel, & Soar, 2016; Ching-Ter, Su, & Hajiyev, 2017).

It has been found that the use of information and communication technologies (ICT) are widely used in educational institutions, especially in higher education institutions. Higher education institutions have invested heavily in the development of online learning systems to support the learning process. The value of investment depends on the implementation of the system used by both students and instructors.(Naveh, Tubin, & Pliskin, 2010; Sharma, Gaur, Saddikuti, & Rastogi, 2017). This is considered as an opportunity for developing online learning (Chang, Hajiyev, & Su, 2017) that reduce time and distance limitations. The instructors and learners can interact in environments that are not traditional ones through electronic learning. The instructors can use the flexibility of the online learning in order to meet the needs of the various grades of students.

The progress of technology-enhanced learning has also allowed higher education institutions to move towards a new dimension in bringing technology to drive learning by developing and creating a learning analytic to analyse students' learning behaviour and patterns for better support. Students can interact directly with the system through the graphic user interface or the dashboard of the system, which offers different features, classified by personalised learning environments Learning analytics help instructors better understand students learning behaviour. (Schumacher & Ifenthaler, 2017). Developing learning analytics is widely available in developed countries (Mavroudi, Giannakos, & Krogstie, 2017; Snodgrass Rangel, Bell, Monroy, & Whitaker, 2015; Wilson, Watson, Thompson, Drew, & Doyle, 2017). Higher education institutions in Thailand have been active in, developing and implementing online learning systems more than the past. The details about online learning in Thailand are described in Sections 1.1 and 1.2, respectively.

**1.1 Background of the Study: Online learning in Thailand**

According to the observation of the uses of ICT by the Southeast Asian Ministers of Education Organization: SWAMEO, it was found that Thailand was considered as a member of the group 2 countries: most developments and uses of ICT were in the infusing step in almost all aspects because the action plans and policies regarding ICT for learning were developed. Nevertheless, it was considered that there were development gaps among the urban and rural areas. Consequently, the developments in some aspects were only in the applying or emerging steps. Moreover, it was found the developments of ICT in Thailand and Vietnam were more significant than those in Indonesia and Philippines (SEAMEO, 2010).

Thai education still focused on improving the education with ICT as a tool for providing the education in order to reduce the learning gaps through online learning. The ministry of Thai education allocated the budgets from creating the technological infrastructures and preparing the instructors' and learners' abilities to use tools. The latest policy encouraged the uses of online learning in higher education institutes for 15 years (2008-2022) by focusing on providing online learning subjects in universities (Commission on Higher Education, 2007). Thai universities also actively changed their traditional teaching method into online learning method. For example, Prince of Songkla University developed the online learning system, "ClassStart", that is available for other universities in Thailand to use this application for free.

*1.2 ClassStart*

ClassStart is an online learning management system developed in order to support Thai educational provision and encourage student-centered learning since users can use it easily and they can access it at any time and place if they just have computers and internet connections. It also supports mobile devices. Class Starts has been free for the instructors and learners all over the country. There are over 45,000 members, over 3,500 classes from over 200 educational institutes in Thailand. The features of Class Start include membership, classroom management, announcement, teaching document, classroom web board, learner group, exercise management, learning record, score, and score processing systems. (ClassStart.org, 2016).

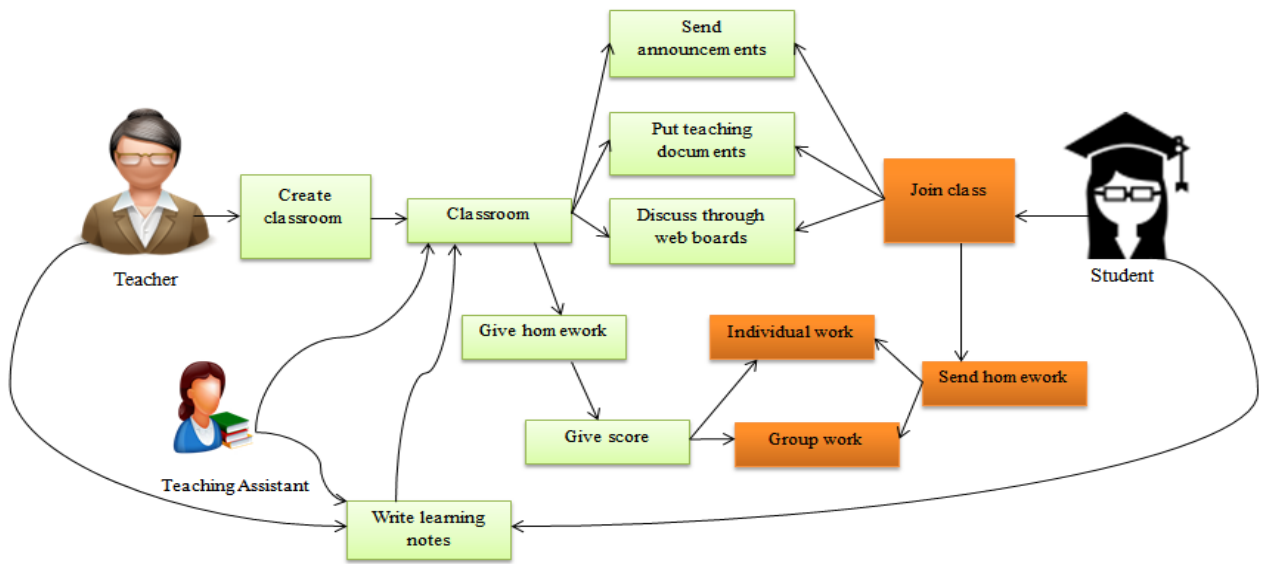

Fig 1. ClassStart Workflow (ClassStart.org, 2016)

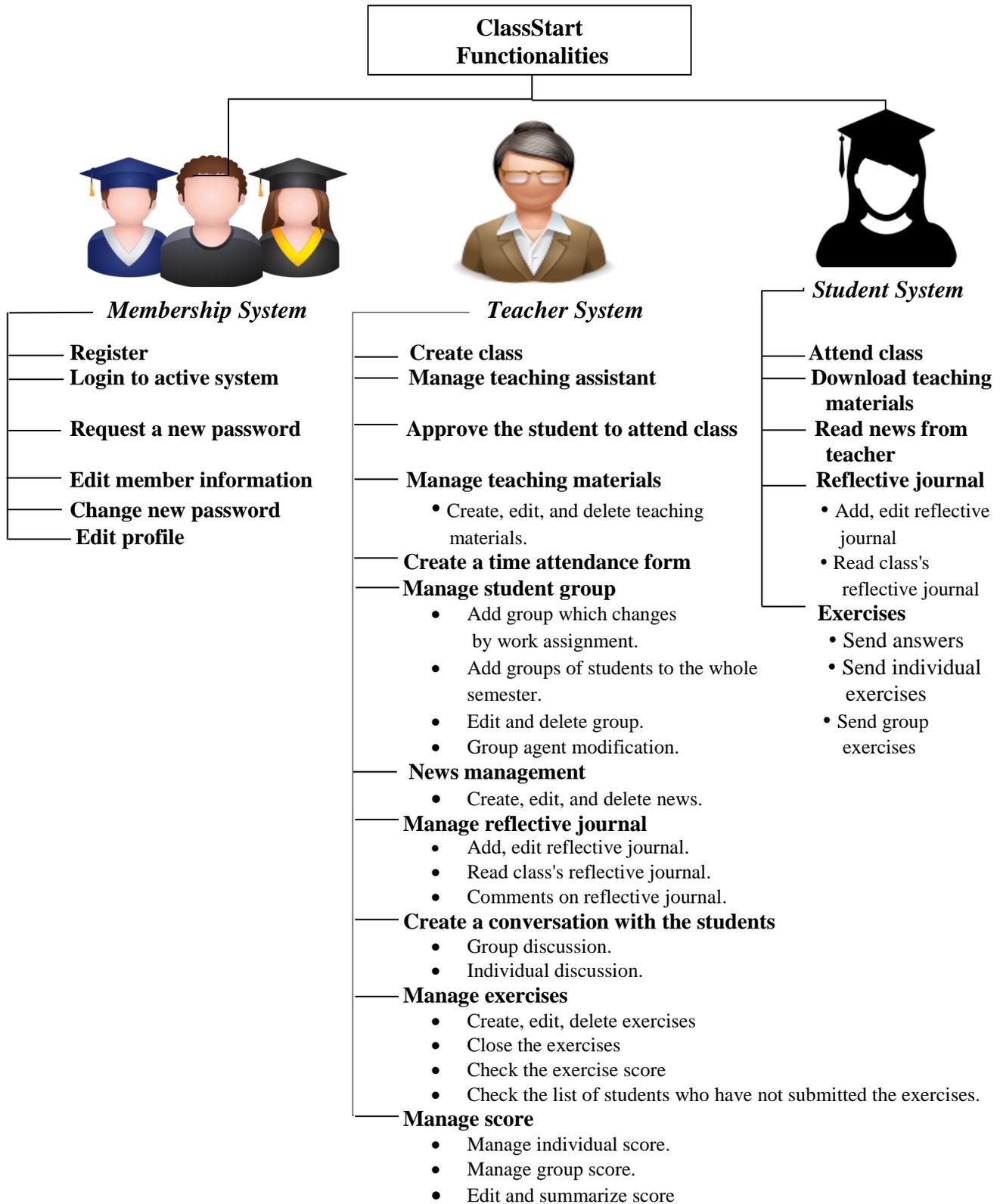

Fig 2. ClassStart Functionalities (ClassStart.org, 2016)

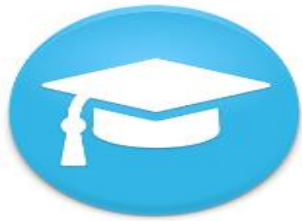
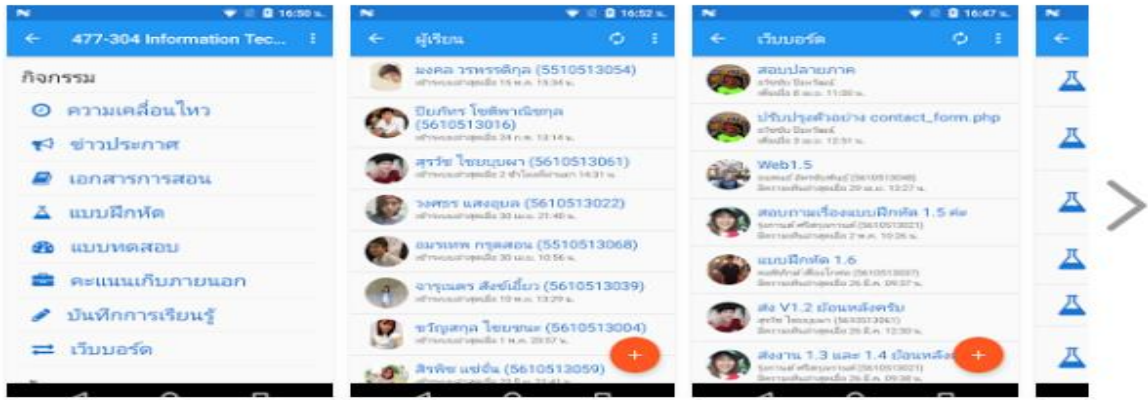

Fig 3. ClassStart (ClassStart.org, 2016)

ClassStart is an educational management system which promotes online learning processes through flexible instruction and technology. It is easy to use with internet access. ClassStart can be divided into three subgroups as membership, teacher, and student systems. The workflow and functionalities of class start shown in figures 2 and 3. The ability of ClassStart enables instructors to apply the ClassStart tools to the learning style of the learners in each class; such as, cooperative learning, active learning, and inductive learning. Moreover, ClassStart offers an individualized approach to personal analytics. For example, a tool like a reflective journal is a record of what has been learned and the learning process that has taken place in the form. Instructors can use this system to track the progress in the students' learning process. Since the procedure of transferring the learning process takes place, it is no less important than the content gained from learning. The transfer of knowledge through regular recording allows learners to connect their learning and learning processes, so that learners can recognize the content and learn more effectively. In addition, it continually monitors student feedback. The instructor can plan the lesson to repair the teaching activities to complete as well.

Although the popularity of the online learning is improving in developing countries such as Thailand, previous studies found that technical problems and the lack of support from relevant individuals resulted in the acknowledgement of the limited value of online learning (Jairak, Praneetpolgrang, & Mekhabunchakij, 2009; James, 2011). Due to previous research found that online learning is popular in developed countries (Chu & Chen, 2016; Duan, He, Feng, Li, & Fu, 2010; Hau & Kang, 2016; Islam, 2013; Liu, Li, & Carlsson, 2010). However, to transfer the technologies from the developed countries to the developing countries, there are different cultures and social characteristics to be considered. Hence, the findings from the developed countries cannot be generalized to the developing ones. The viewpoints from the developing countries are needed to be studied (Avgerou, 2001; Heeks, 2002).

## 1.3 Related works

Online learning adoption has been examined under various contexts in both developing and developed countries. The majority of the previous researchers have studied online learning adoption through psychological variables by applying the concept of technology adoption that is widely used, such as Technology Acceptance Model (TAM) and The Unified Theory of Acceptance and Use of Technology (UTAUT). At the beginning of the study, online learning focused on the acceptance of electronic learning (E-learning), for example Uğur and Turan (2018) studied e-learning acceptance in Turkey in their updated UTAUT model by adding two variables into the research model which are system interactivity and area of scientific expertise.

They point out that the two main variables from UTAUT are performance expectations and effort expectations combined system interactivity contribution to e-learning adoption. In addition, they also found an indirect effect between the area of scientific expertise on behavior and the use of e-learning through performance expectation variables. In line with the result of Dečman (2015) who found that performance expectancy contribute to the acceptance of e-learning in mandatory environments of higher education which applied the UTAUT model together with the two moderators variables (students' previous education and gender). However, the results of the study did not find any element statistically significant in the model fit. Whilst Paola Torres Maldonado, Feroz Khan, Moon, and Jeung Rho (2011) focused on the motivation of learners to use e-learning in developing countries such as Peru by applying the unified theory of acceptance and use of technology (UTAUT) model by adding e-learning motivation variables They found that psychological variables in e-learning motivation and social influence effected students' e-learning attitudes. Moreover, this research also tested the moderating role of region (coast and Andes), it notes that region moderates the relationship between the social influences on the behavior intention to use e-learning. Boateng, Mbrokoh, Boateng, Senyo, and Ansong (2016) studied e-learning adoption in Ghana by applying the concepts of TAM, they point out that perceived usefulness (PU) and attitude towards use (ATTU) had a direct effect on the e-learning acceptance of learners.

Due to a recent boom in mobile learning, there are many researchers focusing on the study of student mobile learning acceptance in higher education. Al-Emran, Elsherif, and Shaalan (2016) said that online learning in the form of mobile learning is a new breath of learning. Ahmad (2013) studied m-learning acceptance among higher education students in Brunel. The results show that the UTAUT model has an effect on learner's intention to use m-learning. They also emphasized that mobile devices moderated the relationship between independent variables including effort expectations, lecturers' influence, quality of service, and personal innovativeness on behavioral intention. With modified TAM, Liu et al. (2010) studied m-learning acceptance of undergraduates students in China, they point out the main variable of TAM, perceived usefulness, is the key factors to determine m-learning adoption. The study is deepened by two aspects: perceived near-term/long-term usefulness. Also, they added the personal innovativeness variable into the research model. The result from statistical analysis confirms the relationship between these three variables on behavior intention. The detail of comparative analysis of online learning acceptance in an academic environment is shown in Table 1, in which the symbols used to confirm that the two variables are statistically correlated they are +, for example (PE + BI).

However, from the literature review, we found that the majority of online learning ignores quality factors; this could be a factor for the success of online learning. Thus, quality studies in all terms are still necessary, especially in terms of learner's opinions towards the system. Another important issue is previous studies lack of in-depth research of specific application in Thailand. Therefore, this research focuses on the ClassStart application. This is an important application in the online learning market of higher education in Thailand.

Table 1. The comparative analysis of online learning acceptance in an academic environment.

| Type of Online Learning | Article | Country | Target group | Adoption Theory | Adoption Determinants | Data analysis Methods Applied | Hypothesis Verification |
|---|---|---|---|---|---|---|---|
| E-learning | Uğur and Turan (2018) | Turkey | Academics | The Modified Unified Theory of Acceptance and Use of Technology (UTAUT) | **IV:** Performance expectancy(PE), Effort expectancy(EE), Social influence(SI), Facilitating condition(FC), System interactivity(SIN) and Scientific expertise (ASE) **DV:** Behavioural Intention | Descriptive statistics, Regression analysis, path' coefficients, significance, and the adjusted coefficients of determination scores | PE + BI SI + BI EE + PE ASE + PE ASE + EE SI + PE ASE + SIN |
| | Dečman (2015) | Slovenia | Students from the Faculty (School) of Administration of the Students | The Unified Theory of Acceptance and Use of Technology (UTAUT) | **IV:** Performance expectancy(PE), Effort expectancy(EE), Social influence(SI) **DV:** Behavioural Intention(BI) | Descriptive statistics, Principal Component Analysis (PCA) | SI + BI PE + BI |
| | Paola Torres Maldonado et al. (2011) | Peru | Students in Peru from different regions which are the coast, Andes and jungle | The Modified Unified Theory of Acceptance and Use of Technology (UTAUT) | **IV:** Social influence(SI), Facilitating conditions(FC), E-learning motivation(ELM), Behavioural intention(BI) **DV:** Use behavior(UB) | Descriptive statistics, Structural equation models(SEM) | ELM+BI BI+UB SI+BI UB+ELM |
| | Boateng et al. (2016) | Ghana | Undergrads students | Technology Acceptance Model (TAM) | **IV:** Perceived ease of use (PEOU), Perceived usefulness (PU), Attitude towards use (ATTU) **EV:** Self-efficacy (CSE) **DV:** E-learning Intention Behaviour (ELIB) | Descriptive statistics, Structural equation models(SEM) | CSE +PEOU PEOU+PU PEOU+ATTU ATTU+ELIB ATTU+ELIB PU+ELIB |
| M-learning | Ahmad (2013) | England | Students from Brunel University. | The Modified Unified Theory of Acceptance and Use of Technology (UTAUT) | **IV:** Performance Expectancy (PE), Effort Expectancy (EE), Lecturers' Influence (LI), Quality of Service (QoS), Personal Innovativeness (PI) **DV:** E-learning Intention Behaviour (ELIB) | Descriptive statistics, Structural equation models(SEM) | PE+BI EE+BI LI+BI Qos+BI PI+BI |
| | Liu et al. (2010) | China | Undergraduate students from Zhejiang Normal University in China | Technology Acceptance Model (TAM) | **IV:** Perceived near-term usefulness (PNTU), perceived long-term usefulness (PLTU), Perceived ease of use (PEOU), Personal innovativeness (PI) **DV:** Behavioural intention (BI) | Descriptive statistics, Convergent validity Discriminant validity Path Analysis | PI+PEOU PI+PLTU PI+BI PLTU+PNTU PLTU+BI PNTU+BI |

**Notes:** IV: Independent Variable, EV: External Variables, and DV: Dependent Variable

Therefore, this study makes three contributions. Firstly, this study intended to increase understanding about the factors affecting online learning in developing countries such as Thailand and reduce the gap in understanding through the new research model by integration psychological construction terms of quality which are: system quality, information quality and service quality from the Delone Mclean IS success model with an individual acceptance of the online learning which are: performance expectancy, effort expectancy and social influence from The Unified Theory of Acceptance and Use of Technology to predict intention to use of online learning. Secondly, the proposed research model tries to study on specific application as Class Start that widely used in Thailand through learners perspective which is directly users. Thirdly, this proposed research model may be used study adoption of the new information system in a various context such as e-commerce, e-health and so forth.

## 2. Theoretical context and hypotheses

*2.1 The Unified Theory of Acceptance and Use of Technology (UTAUT)*

According to the literatures, it was found that the Unified Theory of Acceptance and Use of Technology (UTAUT) is a theory that is broadly used in the field of information and communication technology acceptance. UTAUT includes the models from 8 theories: Theory of Reasoned Action (TRA) from Fred D. Davis (1989), Technology Acceptance Model (TAM) from F. D. Davis (1989) and V. Venkatesh (2000), Motivation Model (MM) from F. D. Davis, Bagozzi, R. P., &Warshaw, P. R. (1992), Theory of Planned Behavior (TPB) from Shirley Taylor (1995), Combined TAM and TPB (C-TAM-TPB) from Shirley Taylor (1995), Model of PC Utilization (MPCU) from Ronald L. Thompson (1991), Innovation Diffusion Theory (IDT) from Gary C. Moore (1991) and Social Cognitive Theory (SCT) from Compeau (1995). Initially, UTAUT focused on the acceptance of technology usage in workplaces. Later, researchers used it to study the acceptance and use of mobile phones as well as web-based applications. It is evident that UTAUT can be applied to the motivation to use technology for education. Dečman (2015) estimate suitability of UTAUT under an e-learning learning environment in higher education. The results found that UTAUT is a prominent model used to study e-learning environments. Milošević, Živković, Manasijević, and Nikolić (2015) examine the effects influence intention to use m-learning of masters students in Serbia. The results show that the performance expectancy variable from UTAUT has the most significant influence on intention. Thus, this result can be used to improve the learning effectiveness and performance of educational institutions. Im, Hong, and Kang (2011) found that the most important part of UTAUT is the relationship between the intention to use and two variables which are: performance expectancy and effort expectancy. A large number of previous research found that performance expectancy is a critical factor in the online learning environment. (Casey & Wilson-Evered, 2012; Gupta, Dasgupta, & Gupta, 2008; Zhou, Lu, & Wang, 2010). According with V. Venkatesh, Morris, M. G., Davis, G. B., & Davis, F. D. (2003) develop his original model which found that performance expectancy was strong predictor of behavior intention in analysis the meta-analysis of UTAUT is 27 cases. As same as meta-analysis of Dwivedi, Rana, Chen, and Williams (2011) founded that performance expectancy was only factor from all 25 cases that had a significant effect on behavior intention. Another important factor that has strong influence of UTAUT is effort expectancy. Most users expect that effort expectations should be low. In context of online learning, researchers found that effort expectancy had influence with experienced student using more than inexperienced students.

The unified theory of acceptance and use of technology (UTAUT) has been widely studied in the context of online learning behavior in Thailand. Jairak et al. (2009) investigated mobile learning adoption by Thai higher education students. Results showed that performance expectancy highly influenced attitudes toward intention to adopt mobile learning. Most Thai students have knowledge and experience regarding online technology; they readily embrace mobile devices as beneficial for learning. Whilst Chularat Saengpassa (2013) stated that the Cyber University program, with the aim to promote and support e-learning at universities, had been operative in Thailand for more than a decade. However, the project has faced obstacles because users as instructors and learners do not fully accept the concept of e-learning. National realization regarding the importance of online learning should be encouraged. Therefore, technology adoption factors applied by the UTAUT theory were examined as part of the research framework. However, UTAUT theory suggests that three variables that determine intentional behavior including: performance expectancy, effort expectancy and social influence (Chiu & Wang, 2008; Oliveira, Faria, Thomas, & Popovič, 2014).

Performance expectancy is the levels that individuals believe that by using the systems, the systems will help him or her achieve a goal (Bao, Xiong, Hu, & Kibelloh, 2013). It involves with perceived usefulness in TAM. Viswanath Venkatesh (2000) found that performance expectancy is an indication factor to predict the usage intention of new technology in operations. In online learning context, students usually find that online learning is useful, convenient, and able to increase study productivity (Wang, Wu, & Wang, 2009). Therefore, it will affect the consistency use of ClassStart. Therefore, we made the hypothesis:

H1: Performance expectancy has a positive effect on behavior intention to use ClassStart.

Effort expectancy is the levels of simplicity and convenience in order to use the system or the outline that learners believe that using the system is free from attempting (Bao et al., 2013). This is because effort expectancy involves with perceived ease of use in TAM. Assuming that any systems that the users realise it is simple to use, thus, there is likely a perception of advantages and intention behavior. In the context of Thailand, Siritongthaworn, Krairit, Dimmitt, and Paul (2006) noted that users were more likely to accept e-learning procedures and processes if they were easy to access and use. However, Information and Communication Technologies (ICT) for teaching and learning often lack service quality and technological problems arise. Relating to Chiu and Wang (2008), it was presented that in a case of effort expectancy increases; it is a lead to a better operation. Thus, effort expectancy should have a direct impact on intention. Therefore, we made the hypothesis:

H2: Effort expectancy has a positive effect on behavior intention to use ClassStart.

Social influence is personal levels of recognition that people who are close or important to them believe that they should use technology or new systems. Social influence pertains to subjective norm in theory of planned behavior (TPB). It found that users tend to communicate with others to outline their technology acceptance (Dečman, 2015; Magsamen-Conrad, Upadhyaya, Joa, & Dowd, 2015). The previous study showed that subjective is an important factor that indicates system usage intention relating to a research done by Milošević et al. (2015). The researcher found that an influence from surrounded people such as instructors has an impact on the usage intention of online learning by learners. Therefore, we made the hypothesis:

H3: Social influence has a positive effect on behavior intention to use ClassStart.

*2.2 Delone and McLean IS success model*

System design and a system with respectable functionality is one of the cornerstones of system development life cycle. It is also key factor to ensuring the success of the implementation of the information system (W. H. DeLone & McLean, 1992; Detlor, Hupfer, Ruhi, & Zhao, 2013). Learners are more likely to use online learning if they find that the online learning system has quality. (H.-F. Lin, 2007). The quality of the application is important. Especially, educational application because online learning application lack of face to face between learner and instructor. With online learning context, learners use system for learning activities. The updated D&M IS success model can be used to study online learning system (Wang, Wang, & Shee, 2007).

Initially, IS success model was designed by W. H. DeLone and McLean (1992) to evaluate information system success. This model presents six key success variables – (1) system quality, (2) information quality, (3) information system use, (4) user satisfaction, (5) individual impact and (6) organization impact. Later, William H Delone and McLean (2003) presented updated IS success model and assessed benefits arising from IS practice changes, especially increasing Internet use. Thus, William H Delone and McLean (2003) added "service quality" as a new dimension in IS success model. Also, "impact" was become "net benefits". Thus, updated model including 6 variables: (1) information quality, (2) system quality, (3) service quality, (4) use/intention to use, (5) user satisfaction, and (6) net benefits. The majority of previous research papers applied IS success model, presented by William H Delone and McLean (2003) for online learning development and assessment (Wang et al., 2007). William H Delone and McLean (2003) presents that updated IS success model can adjust to fit online learning's challenge assessment.

According to a study by the Thai Ministry of Education, the country is still struggling to deliver efficient information systems in terms of service quality, adequate technical backup and support, and available resources/courseware developers (Rueangprathum, Philuek, & Fung, 2009). However, researchers continue to focus on the quality of e-learning in Thailand. Wannapa Khaopa (2012) recognized Thailand as the online learning hub of the ASEAN region and the country has many cosmopolitan courses and programs which attract international students, especially from ASEAN countries. Online learning is a new electronic learning market that allows students to study at home. In addition, Pruengkarn, Praneetpolgrang, and Srivihok (2005) studied the quality of e-learning at Thai higher education institutions according to the Institute of Electrical and Electronics Engineers

(IEEE) standards. They determined the average quality of e-learning websites in Thailand in terms of functionality, reliability, usability, efficiency, maintainability and portability at 50.34%, which rates as moderately aggressive. Thus, lack of service quality is hampering the use of ICT for teaching and learning. Consequently, this research takes three quality variables in update IS success - system quality, information quality and service quality to study on intention to use ClassStart.

System quality refer to the quality of the information system that user can use the system easily, as the availability of the system, speed of response (feedback), user-friendly and features of the screen (Interface), which is the quality of the information system will affect the intention to use the system (William H Delone & McLean, 2003). Poor system quality may result in student dissatisfaction. If the learner finds that the system is difficult to use. This may result in the intention to make the system to fail. The previous researchers found that system quality influences intention to use the information system such as Lee and Hsieh (2009) found that the system quality affects directly to the use of mobile data services. In line with McFarland and Hamilton (2006) showed that system quality were affected on perceived use and actual use the system. Therefore, we made the hypothesis:

H4: System quality has a positive effect on behavior intention to use ClassStart.

Information quality refer to quality of data that user received when used information system, as relevant, easy to understand, up to date and completeness. Whereof quality of information will affect intention to use and user satisfaction of the system (William H Delone & McLean, 2003). Poor information quality may affect the reliability of the system for learning. This reduces the impulse to use the system. Consequently, Information quality is an important and required for education. Especially for online classes, such as e-learning or m-learning that information quality can be a success factor of the system (Williams & Jacobs, 2004). Thus, the assessment of information quality is important dimension of quality online learning (Byrd, Thrasher, Lang, & Davidson, 2006). Therefore, we made the hypothesis:

H5: Information quality has a positive effect on behavior intention to use ClassStart.

Service quality is the quality that user receives help or answer questions related to the issue of information systems from the provider system as sincere interest in solving problem, personalization, trust and understand user specific need. And the quality of service will affect in the intended use of the system (William H Delone & McLean, 2003). Poor service quality, such as when students have problems using the system, whereby the system is delayed or does not meet their requirements. This may result in decreased motivation for the system. A large number of previous studies focused on service quality. In education context found that students are ready to take online learning applications if they perceived that have the quality of service. That is consistent with Milošević et al. (2015) found that service quality has a direct impact on intention to use m-learning of student in Serbia. E. Park and Kim (2013) found that service quality has a direct impact on acceptance the long-term evolution (LTE) service of students in South Korea Therefore, we made the hypothesis:

H6: Service quality has a positive effect on behavior intention to use ClassStart.

Therefore, Fig. 4 shows our model; in addition to the core constructs of UTAUT and IS success model were assumed to affect the intention to use ClassStart.

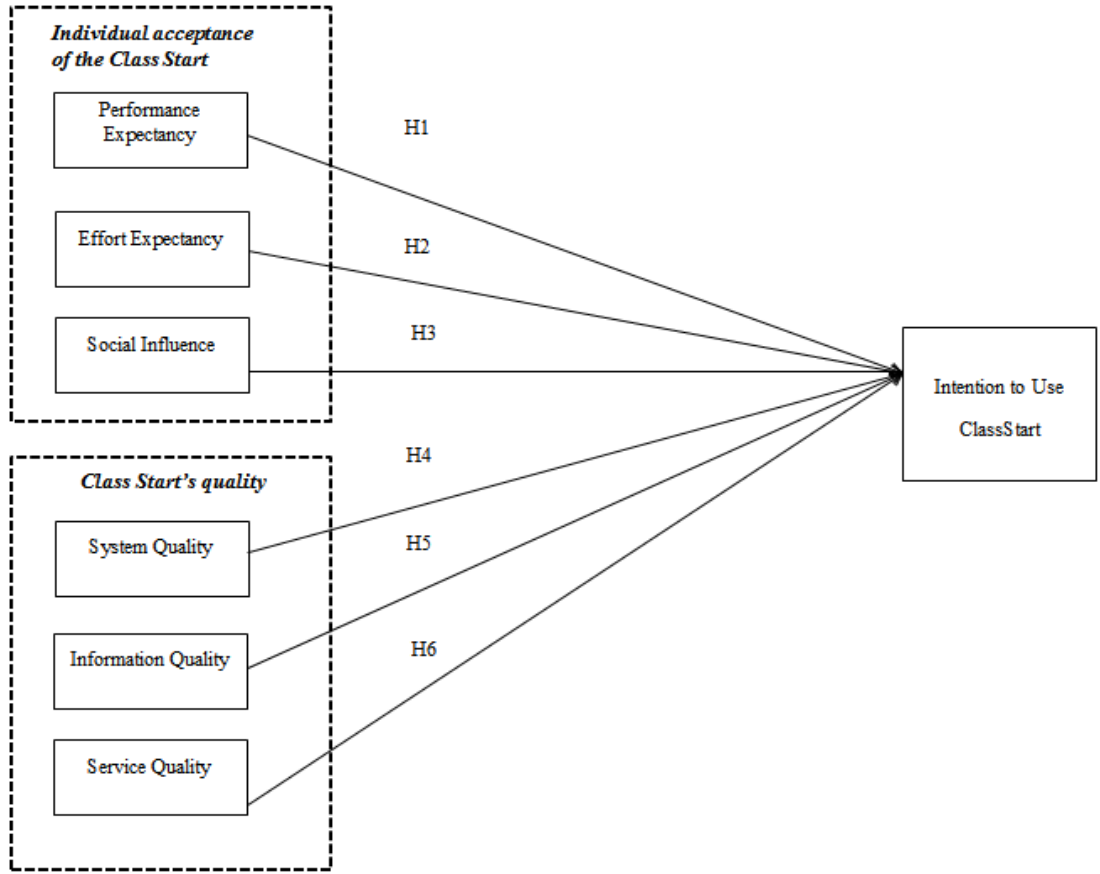

Fig 4. The research model.

## 4. Research methodology

### *4.1 Instrument*

This research consists of seven constructs from two dimensions: Class Start App's qualities which consist of system quality, information quality, and service quality and individual acceptance of the ClassStart which consist of performance expectancy, effort expectancy and social influence. All of the variables in this research are based on previous researches.

First, research questions were developed in English based on a literature review. Second, the research questions were translated into Thai. The researcher contacted two experts in the field of information systems and two from linguistic fields to considering the research questions. Third, based on their suggestion, the researchers adjusted some items. After the questionnaire was completed the researcher sent a pre-test to 30 students (female 20, male 10) who have had experience of the ClassStart. Using their suggestions the researchers modified the questionnaire to ensure that it was unambiguous and ensuring clear understanding. The finished questionnaire is shown in Appendix A. Each question uses the Likert scale measurements from "strongly disagree" (1) to "strongly agree" (5).

*4.2 Procedures*

As ClassStart is the largest online learning provider in Thailand, the questionnaires were distributed to universities that have participating in using the ClassStart. This study took 3 months to distribute the questionnaire in person. The non-probability convenience sampling method was used because of the limited time and resources (Sekaran & Bougie, 2011). After the removal of incomplete returned questionnaires. There were 307 completed questionnaires.

*4.3 Participants*

The respondents were dominated by females 230 (74.9%) while males were 77 (25.1%). The majority of the respondents were between 22-25 years of age (60.1%). The demographic information of respondents is show in Table 2.

**Table 2** Demographic profiles and descriptive statistics of respondents. (N = 307).

| Characteristics | Frequency | Percent |
|---|---|---|
| *Gender* | | |
| Male | 77 | 25.1 |
| Female | 230 | 74.9 |
| *Age* | | |
| 18–21 | 117 | 32.6 |
| 22–25 | 216 | 60.1 |
| 26–30 | 26 | 7.2 |
| *Faculty* | | |
| Science | 148 | 48.2 |
| Management | 125 | 40.7 |
| Humanities and Social Sciences | 32 | 10.4 |
| Other | 2 | 0.7 |

## 5. Data analysis and results

### 5.1 Partial least square analysis methodology

Partial Least Squares (PLS) regression analysis is suitable: 1. To predict or test theoretical models that have been studied and proposed in various contexts and 2. To identify or test the relationships between variable (Sosik, Kahai, & Piovoso, 2009). Therefore, SmartPLS 2.0.M3 was used in this study to predict factors that affect the intention to use ClassStart. Statistics used in the research is Structural Equation Modelling (SEM) based on two step approaches introduced by Anderson and Garbing (1988). First, we analysed the measurement model to test reliability and validity, and second, we analysed structural model to test hypotheses.

### 5.2 The measurement model

The measurement model analysis including: internal reliability, convergent validity and discriminant validity (Joseph F. Hair, Anderson, Tatham, & Black, 1995). The internal reliability was examined by considered Cronbach's alpha and composite reliability (CR) greater than 0.7. Convergent validity was examined by considered will be accepted when measurement constructs and item loading are greater than 0.5. Table 3 show Cronbach's alpha value range from 0.80 to 0.85, composite reliability (CR) value range from 0.84-0.90, and item loading value range from 0.75 to 0.89. All Cronbach's alpha, composite reliability (CR) and item loading exceeded recommended value. The above results indicate that the conditions for internal reliability and convergent validity are satisfactory for this study. In parts of discriminant validity was examined by square root of AVE and cross loading matrix. The satisfactory discriminant validity must have square root of AVE was larger than its correlation with other factors (Henseler, Ringle, & Sinkovics, 2009). Table 4 show bold diagonal show square root were greater than corresponding correlation. Thus data in this study had good discriminant.

Table 3. Measurement model.

| Constructs | Items | Loadings | CR | Cronbach's alpha | AVE |
|---|---|---|---|---|---|
| Behavioral Intention | BI1 | 0.75 | 0.90 | 0.85 | 0.70 |
| | BI2 | 0.86 | | | |
| | BI3 | 0.86 | | | |
| | BI4 | 0.84 | | | |
| System Quality | STQ1 | 0.83 | 0.89 | 0.82 | 0.73 |
| | STQ2 | 0.89 | | | |
| | STQ3 | 0.85 | | | |
| | STQ4 | 0.84 | | | |
| Information Quality | INTQ1 | 0.80 | 0.89 | 0.81 | 0.72 |
| | INTQ2 | 0.88 | | | |
| | INTQ3 | 0.84 | | | |
| | INTQ4 | 0.86 | | | |
| Service Quality | SERQ1 | 0.80 | 0.89 | 0.81 | 0.72 |
| | SERQ2 | 0.87 | | | |
| | SERQ3 | 0.85 | | | |
| | SERQ4 | 0.88 | | | |
| Effort Expectancy | EE1 | 0.78 | 0.84 | 0.82 | 0.68 |
| | EE2 | 0.82 | | | |
| | EE3 | 0.85 | | | |
| Performance Expectancy | PE1 | 0.75 | 0.90 | 0.84 | 0.68 |
| | PE2 | 0.88 | | | |
| | PE3 | 0.87 | | | |
| | PE4 | 0.80 | | | |
| Social Influence | SI1 | 0.80 | 0.88 | 0.80 | 0.71 |
| | SI2 | 0.85 | | | |
| | SI3 | 0.87 | | | |
| | SI4 | 0.85 | | | |

Table 4. Correlation Matrix and Square Root of the Average Variance Extracted.

|      | EE      | BI      | INTQ    | PE      | SI      | SERQ    | STQ    |
|------|---------|---------|---------|---------|---------|---------|--------|
| EE   | **0.82462** |         |         |         |         |         |        |
| BI   | 0.4722  | **0.83666** |         |         |         |         |        |
| INTQ | 0.6096  | 0.5547  | **0.84853** |         |         |         |        |
| PE   | 0.6758  | 0.6149  | 0.6115  | **0.82462** |         |         |        |
| SI   | 0.5642  | 0.5564  | 0.5371  | 0.6629  | **0.84261** |         |        |
| SERQ | 0.583   | 0.5098  | 0.677   | 0.5559  | 0.4806  | **0.84853** |        |
| STQ  | 0.5519  | 0.4813  | 0.6344  | 0.5759  | 0.4879  | 0.6434  | **0.8544** |

### *5.3 The structural model*

The structural model was created to identify the relationship between factors in research model. Bootstrap method was conducted to test research hypotheses (Efron & Tibshirani, 1994). Table 5 show detail results (Path coefficient [β and t statistics]). The research model accounts for behavioral intention to use ClassStart (BI) = 58.1%. It was found that system quality (t= 2.122, β = 0.038), information quality (t= 2.294, β = 0.173), performance expectancy (t= 3.262, β = 0.320) and social influence (t= 3.066, β = 0.205) had significant effect on intention to use ClassStart. Thus, H1, H2, H3, H4 and H6 were supported. While service quality (t= 1.527, β = 0.122) and effort expectancy (t= 0.796, β = -0.050) had no significant effect on intention to use ClassStart. Thus, H3 and H5 were not supported.

Table 5. Structural model.

| Path | Coefficient (β) | t-Statistics | Comments |
|---|---|---|---|
| STQ -> BI | 0.038 | 2.122 | Supported |
| INTQ -> BI | 0.173 | 2.294 | Supported |
| SERQ -> BI | 0.122 | 1.527 | Not Supported |
| PE -> BI | 0.320 | 3.262 | Supported |
| EE -> BI | -0.050 | 0.796 | Not Supported |
| SI -> BI | 0.205 | 3.066 | Supported |

Note. P-Value <0.05

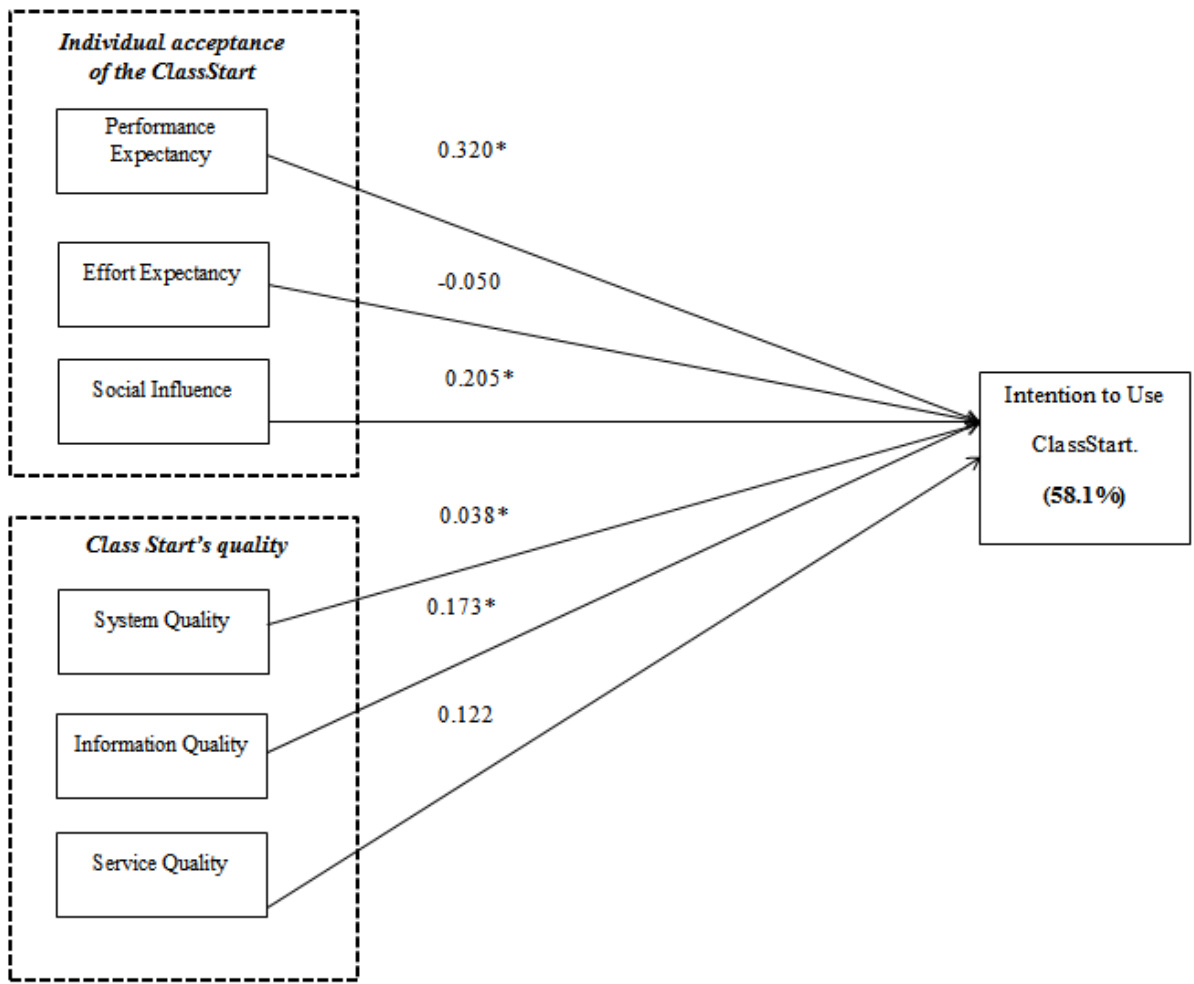

Note. * $p<0.05$; $R^2$ values are shown in parentheses
Fig 5. Standardized path coefficients.

## 6. Discussion and Conclusions

### *6.1 General Discussion and Theoretical Implications*

The research model presented integration constructs that show the Class Start's quality, which was adopted according to the IS Success model and the individual acceptance of the Class Start App as adopted by UTAUT theory. Both dimensions can explain the intention to use the Class Start ($R^2$ = 58.1%).

The results of this study are consistent with previous research, which confirmed that performance expectancy was the most important variable to predict the intention to use technology (Casey & Wilson-Evered, 2012; Gupta et al., 2008; Zhou et al., 2010). In the context of online learning, Ahmad (2013) believes that students with high performance expectations are more likely to accept online learning than those with low expected performance. This is because learners believe that online learning will help them to achieve their expected success as well as improve the efficiency of their learning.

Effort Expectancy is another important factor to study the intention to use online learning (Marchewka, Liu, & Kostiwa, 2007; Murali, 2012) However, this study found that effort expectancy had a negative influence on the intention to use the ClassStart. The results are in line with Milošević et al. (2015), who studied the intention to use online learning in developing countries such as Serbia. This may be because the adoption of technology in developing countries is still in its relative infancy. Learners may not be familiar with changing the traditional learning approach to online learning. The students don't feel flexibility and it takes a lot of effort to use the system.

Social influence also affects the intention to use the ClassStart. Rogers (1995) found that interacting with people in the surrounding vicinity influenced the interpretation of IT adoption. Because students often communicate with others such as classmates and teachers, Lu, Yao, and Yu

(2005) found that a lecturer's influence was an important issue in learners' acceptance of online learning.

Findings concurred with (Jairak et al., 2009; Ngampornchai & Adams, 2016) and research in developing countries such as Ghana (Boateng et al., 2016) and China (Liu et al., 2010); learners recognized that the benefits of e-learning affected their acceptance of this novel educational method.

From the discussion above, we offer the user's perspective on technology adoption. Another issue highlighted in the research is the quality of the ClassStart. This research found that information quality and system quality had some influence on the intention to use the ClassStart. The results of this study are consistent with previous research, which confirmed that system and information quality were important indicators for the adoption of user technology (Gan & Balakrishnan, 2016). In developing countries such as Oman, Al-Busaidi (2013) and Sharma et al. (2017) found that the main issue affecting the acceptance of learners was system quality. Therefore, a successful e-learning system must offer high functionality with exceptional service quality.

Information quality had the most influence on the intention to use the ClassStart. This is because online learning application programs act as a mediator between the learner and the instructor. This finding is consistent with previous research confirming that information quality was an important variable that affected the intention to use technology (Ahn, Ryu, & Han, 2007; Alla & Faryadi, 2013; William H Delone & McLean, 2003). Possibly because the learner realizes the application presents information that is relevant, timely and easy to understand, which affects the decision-making process to use ClassStart.

Aside from the previously reported information, this research found that system quality also influenced the intention to use the ClassStart. This result is consistent with previous research that confirmed that the user's decision to use the system utilize will be affected by awareness of system attributes such as high quality, user-friendly accessibility and interactivity promoting ease of use (Gan & Balakrishnan, 2016).

It was surprising that service quality had little or no influence on the intention to use the ClassStart since previous research appeared to confirm that service quality was an important element in the success of information systems. Due to service quality meeting the needs of users in the adoption of information systems, service quality for online learning is an important part of the educational process (Hassanzadeh, Kanaani, & Elahi, 2012; Milošević et al., 2015). Moreover, the devices used for online learning also vary such as personal computers, computer networks and mobile devices (Milošević et al., 2015). Students will use online learning when the quality of the system is perceived to be beneficial to the learner(Chong, Chong, Ooi, & Lin, 2011; S. Y. Park, Nam, & Cha, 2012). However, this study did not focus on the influence of service quality on the intention to use the ClassStart. It may be that ClassStart has not been applied to all subjects, while other subjects are not used continuously. Thus, students have not been through this issue important to the decision to use ClassStart.

From theoretical implications, this research applies Delone and McLean IS success model and The Unified Theory of Acceptance and Use of Technology to describe the intention to use the ClassStart. We found that the applied variables of two theories could better explain the intention to use with regard to performance expectancy, social influence, information quality and system quality. Thus, study about information systems, especially the education system, should not be concerned only about the user's perspective on technology adoption based on UTAUT and TAM, but should pay attention to the quality of the system. Although users have a positive attitude toward an acceptable information system, they still have problems when using the system, especially quality problems. William H Delone and McLean (2003) stated that when the information system is not available in quality aspect such as an inadequate of the service system and information incorrectness. These may affect learners' acceptance. In the educational system, quality factors play the important role in teaching and learning. Importantly, it also resulted in learner's intention (Kibelloh & Bao, 2014b). The results presented in this research can be used as the guideline in the context of developing countries like Thailand which is facing with these problems. Thus, future research can apply this research framework to study online learning adoption in various contexts such as mobile learning and web-based learning.

*6.2 Practical implications*

From practical implications, our research shows that performance expectancy had the most significant effect on intention to use ClassStart. We also found the influence of information and service qualities on intention to use the ClassStart. Thus, we suggest developers and designers of web-based learning sites concerned with the development of information systems should be aware of the quality of mobile applications because system quality and information quality are predictors of acceptance. It should also be important in the process of analysis, design and implementation as well as developing a reliable, flexible and timely system. By the way, the accuracy of input and output is important and should not be overlooked along with interface design to make it interesting. In addition, educational institutions, educational managers and teachers must continue to push for continuous learning through online learning to make students more familiar with studying online. Also, promoting learners' perspective in using online learning can achieve academic success by offering ease of use and flexibility, which reduces effort expectancy and increases performance expectations.

*6.3 Managerial implications*

From managerial implications, the Ministry of Education, education adminstrator and related agencies should establish a master plan to enhance the use of technology-enhanced learning in the development of educational resources in various areas.

- Improve the learning quality of learners, encouraging learners to take advantage of ICT for learning from a variety of sources and methods.

- Develop the quality of instructors, produce and develop personnel to meet the demand for ICT. Encourage training to keep up with the advancement of technology and bring the instructors into a sustainable learning environment.

- Encourage the development of online learning and development of learning analytical tools to analyse and develop an online learning system to meet the needs of future learners as well as develop the quality of online learning in terms of information, systems and services, as well as develop ICT infrastructure to expand the access to education services to the fullest.

*6.4 Limitations and Future Directions*

There are interesting limitations in this study, First of all, we describe the adoption of the ClassStart user through the UTAUT and Delone and McLean IS success model. However, there may be other causes that affect future intention for use. The research may lead to other factors such as computer self-efficacy and perceived mobility. Secondly, we focused on the ClassStart. The sample used in this research was the learner. Future studies should expand on the current research to include the use or focus on other applications or focus on the learning analytical system to bring the data to be analysed and demonstrate the development or learning outcomes of the students. Thirdly, user behavior is dynamic and constantly changing. Long-term research may provide insights into how users will accept use over time. Finally, we carried out research in Thailand, where online learning is still in its early stages. Our results may not even cover other countries, where online learning is more widely available. Future research can test against online learning acceptance in other countries.

*6.5 Conclusion*

Due to the ability of technology to enhance learning, many countries have become more aware of technology used to promote education, including developing countries like Thailand. This has changed the focus of student-centered instruction from limited learning in the classroom to learning through the internet. By integrating UTAUT and Delone and McLean IS success model, this research analyzes learner adoption of educational applications. Our results show that learners' technology adoption as well as the quality of the application impact overall adoption.

# Appendix
## Appendix A. Scales

**Performance Expectancy (PE)** (adapted from (Chiu & Wang, 2008; Milošević et al., 2015; V. Venkatesh, Morris, M. G., Davis, G. B., & Davis, F. D., 2003))

**PE1:** I find ClassStart useful for my studies.
**PE2:** Using ClassStart would enable me to achieve learning tasks more quickly.
**PE3:** Using ClassStart in my studying would increase my learning productivity.
**PE4:** ClassStart could improve my collaboration with classmates.

**Effort Expectancy (EE)** (adapted from (Chiu & Wang, 2008; Milošević et al., 2015; V. Venkatesh, Morris, M. G., Davis, G. B., & Davis, F. D., 2003))

**EE1:** I would find ClassStart flexible and easy to use.
**EE2:** Learning to operate ClassStart does not require much effort.
**EE3:** My interaction with ClassStart would be clear and understandable

**Social Influence (SI)** (adapted from (Chiu & Wang, 2008; Milošević et al., 2015; V. Venkatesh, Morris, M. G., Davis, G. B., & Davis, F. D., 2003))

**SI1:** I would use ClassStart if it was recommended to me by my lecturers.
**SI2:** I would use ClassStart if it was recommended to me by my classmate.
**SI3:** I would like to use ClassStart if my lecturers' supported the use of it.
**SI4:** People who are important to me think that I should use ClassStart.

**System Quality (STQ)** (adapted from (Ahn et al., 2007; Al-Busaidi, 2012; H.-H. Lin, Wang, Li, Shih, & Lin, 2016) )

**STQ1:** ClassStart system provides high availability.
**STQ2:** The response time of the ClassStart system is reasonable.
**STQ3:** ClassStart system has attractive features to appeal to the users
**STQ4:** ClassStart system provides interactive communication between teacher and students

**Information Quality (INTQ)** (adapted from (Ahn et al., 2007; Al-Busaidi, 2012; H.-H. Lin et al., 2016) )

**INTQ1:** The information provided by ClassStart is relevant.
**INTQ2:** The information provided by ClassStart is easy to understand.
**INTQ3:** The information provided by ClassStart is up to date.
**INTQ4:** The information provided by ClassStart is complete.

**Service Quality (SERQ)** (adapted from (Ahn et al., 2007; Al-Busaidi, 2012; H.-H. Lin, Wang, Li, Shih, & Lin, 2016))

**SERQ1:** When you have a problem, the ClassStart service shows a sincere interest in solving it.
**SERQ2:** ClassStart service is always willing to help you.
**SERQ3:** ClassStart service gives you individual attention.
**SERQ4:** ClassStart service understands your specific needs.

**Intention to use ClassStart (BI)** (adapted from (Cheng, 2015; Liu et al., 2010; Milošević et al., 2015))

**INT1:** I intend to use ClassStart device for learning in the future.
**INT2:** I predict that I will use ClassStart frequently.
**INT3:** I will enjoy using ClassStart. .
**INT4:** I would recommend others to use ClassStart.